\documentclass[twocolumn,english]{revtex4}
\usepackage[T1]{fontenc}
\usepackage[latin9]{inputenc}
\usepackage{amstext}
\usepackage{graphicx}

\makeatletter
\@ifundefined{textcolor}{}
{%
 \definecolor{BLACK}{gray}{0}
 \definecolor{WHITE}{gray}{1}
 \definecolor{RED}{rgb}{1,0,0}
 \definecolor{GREEN}{rgb}{0,1,0}
 \definecolor{BLUE}{rgb}{0,0,1}
 \definecolor{CYAN}{cmyk}{1,0,0,0}
 \definecolor{MAGENTA}{cmyk}{0,1,0,0}
 \definecolor{YELLOW}{cmyk}{0,0,1,0}
 }

\makeatother

\usepackage{babel}

\makeatother

\usepackage{babel}

\makeatother

\usepackage{babel}

\makeatother

\usepackage{babel}

\begin{document}

\title{Time domain detection of pulsed spin torque damping reduction}

\author{Longfei Ye$^{1}$}

\author{Samir Garzon$^{1}$}

\author{Richard A. Webb$^{1}$}

\author{Mark Covington$^{2}$}

\author{Shehzaad Kaka$^{2}$}

\author{Thomas M. Crawford$^{1}$}

\affiliation{$^{1}$Department of Physics and Astronomy and USC Nanocenter, University
of South Carolina, Columbia, SC 29208, USA.\\
 $^{2}$Seagate Research, 1251 Waterfront Place, Pittsburgh, PA
15222, USA.}
\begin{abstract}
Combining multiple ultrafast spin torque impulses with a 5 nanosecond
duration pulse for damping reduction, we observe time-domain
precession which evolves from an initial 1 ns duration transient with
changing precessional amplitude to constant amplitude oscillations
persisting for over 2 ns. These results are consistent with relaxation of the transient trajectories to a stable orbit with nearly zero damping. We find that in order to observe complete damping cancellation and the transient behavior in a time domain sampling measurement, a short duration, fast rise-time pulse is required
to cancel damping without significant trajectory dephasing. 
\end{abstract}
\maketitle
Spin-polarized electrons passing through a nanoscale ferromagnet~(nanomagnet)
exert a spin-transfer torque~(STT)~\cite{slonczewski_JMMM1996,berger_PRB1996}
on the local magnetization. In contrast to external field torques,
STT can be largely collinear with the Landau Lifshiftz Gilbert damping
torque~\cite{ralph_JMMM2007}, allowing control over damping~\cite{krivorotov_SCIENCE2005,sankey_PRL2006},
in addition to driving precession and switching~\cite{tsoi_PRL1998,katine_PRL2000}.
Spin torque's ability to coherently cancel the damping torque holds
great potential for applications of nanopillars in high stability,
low-linewidth microwave oscillators~\cite{katine_JMMM2008}. Whereas for ac driven oscillators the linewidth depends directly on damping, linewidths of dc driven oscillators are dominated by trajectory dephasing~\cite{sankey_PRB2005,sankey_PRL2006}. Measuring and understanding the effects of damping and dephasing is
critical for minimizing oscillator linewidths and maximizing their
power output~\cite{houssameddine_PRL2009}. At low temperatures,
time-domain~\cite{krivorotov_SCIENCE2005} and ferromagnetic resonance
(FMR) measurements~\cite{sankey_PRL2006} of spin-valve nanopillars
have demonstrated spin torque induced damping reduction. Here, we
demonstrate damping cancellation at room temperature by using a pulsed
time domain technique: two ultrafast (30 ps) spin torque impulses
excite nanomagnet dynamics and the switching probability~$P_{S}$
is measured as a function of relative pulse-pulse delay~\cite{garzon_PRB2008}.
Our measurement maps a precessional trajectory to a digital response,
either switching or nonswitching. We compare damping cancellation
using a 5 ns current pulse and a dc current, and show that the dc
current leads to switching and dephasing, preventing complete cancellation
of damping. However, the 5 ns pulse, at amplitudes well above the
{}``dc'' critical current, can cancel the effective damping and
yield coherent dynamics over our maximum time window of 2.2 ns. Our
technique allows us to observe the transient dynamics that precede
orbit stabilization in the presence of a constant current, and the
trajectory dephasing that occurs over a longer timescale. 


%
\begin{figure}[t]

\begin{centering}
\includegraphics[width=3.3in]{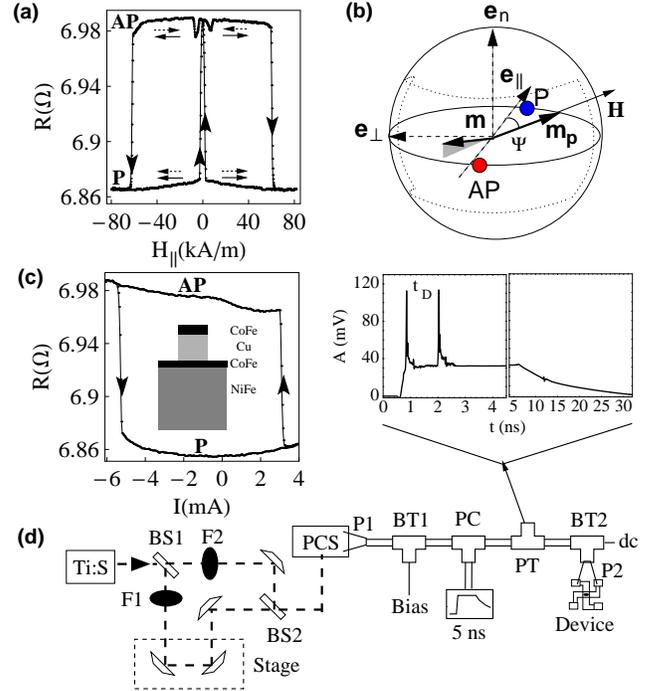}\\
 
\par\end{centering}

\caption{\textbf{Device characterization and experimental setup.} (a) Resistance
vs easy-axis magnetic field, $H_{\parallel}$, for a typical device
at room temperature. (b) Schematic describing the orientation of the
polarizer $\mathbf{m_{p}}$ and P, AP stable points. $\mathbf{m_{p}}$
tracks the applied field, $H$. (c) Resistance vs. dc current for
device shown in (a) with H=46 kA/m and $\Psi\sim$12 degrees. Inset:
device schematic. (d) Experimental setup for time-resolved measurements
of spin torque switching with damping cancellation. Inset: voltage
waveform (ultrafast pulse pair plus 5 ns pulse) measured at a pick-off
tee before the device.}

\centering{}\label{fig:fig1} 
\end{figure}

Spin-valve nanopillar devices with elliptical cross section were fabricated
using e-beam lithography and ion milling. The devices are composed
of an extended Ni$_{80}$Fe$_{20}$(20 nm)/Co$_{90}$Fe$_{10}$(2
nm) polarizer which provides a large magnetic moment to improve its
stability, a 10 nm Cu spacer, and a 3 nm Co$_{90}$Fe$_{10}$ {}``free''
layer or nanomagnet {[}Inset to Fig.~\ref{fig:fig1}(c){]}. A typical
plot of resistance vs. field is shown in Fig.~\ref{fig:fig1}(a),
where the large vertical arrows indicate free layer switching between
P and AP states (largely parallel or anti-parallel to the polarizer).
The small dashed and solid arrows represent the orientation of the
free layer, $\mathbf{m}$, and polarizer, $\mathbf{m_{p}}$, respectively.
The measured Stoner-Wohlfarth astroid for our devices indicates that
the polarizer follows the applied field, reversing orientation around
zero field {[}small dips in resistance in Fig.~\ref{fig:fig1}(a){]}.
Polarizer reversal near zero field, along with considerable interlayer
dipole coupling, are responsible for the two symmetric loops at positive
and negative fields. For all STT measurements we apply an in-plane
magnetic field $H$ at a small angle $\Psi\sim$10 degrees with respect
to the easy axis $\mathbf{e_{\|}}$ {[}Fig.~\ref{fig:fig1}(b){]}
to cancel the polarizer's dipolar field, facilitate AP-P current induced
switching, and set the orientation of the polarizer to obtain a non-collinear
geometry, increasing switching reproducibility~\cite{garzon_PRB2008}.
For the values of $H$ used throughout our experiments the orientations
of points P and AP are displaced less than 3 degrees from $\mathbf{e_{\|}}$
due to the easy axis anisotropy of the free layer. Current, defined
as positive when flowing across the multilayer from top to bottom
{[}inset to Fig.~\ref{fig:fig1}(c){]} can be used to switch the
nanomagnet via STT~\cite{ralph_JMMM2007}. A typical resistance vs
dc current loop {[}Fig.~\ref{fig:fig1}(c){]} shows sharp transitions
between the P and AP states and a $\sim$8 mA wide region of bistability.
Four terminal measurements of the P state resistance and magnetoresistance
give typical values $\sim$1.6$\Omega$ and $\sim$7$\%$ respectively.
The large two terminal resistance values in Figs. \ref{fig:fig1}(a),
(c) are due to lead resistance and inductive impedance of the bias
tees shown in \ref{fig:fig1}(d).

To generate ultrafast spin torque current pulses we use an amplified
Ti:Sapphire mode-locked laser (120fs FWHM,~$\sim$1.6mJ per pulse)
in single-shot mode {[}Fig. \ref{fig:fig1}(d){]}. An optical pulse
is split into two separate pulses at the first beamsplitter (BS1),
with each pulse having independently controlled amplitudes via tunable
optical filters (F1, F2). A sub ps resolution variable optical delay
$t_{D}$ between the pulses is produced with a translation stage,
after which the beams are recombined at BS2 and focused onto a Au
photoconductive switch~(PCS)~\cite{auston_1984}. The two PCS generated
electrical pulses (14.9 mA, $\sim$ 30 ps FWHM) are sent through a
40 GHz coplanar-to-coaxial probe (P1) to the PCS bias tee (BT1) with
12 ps risetime and 40V dc switch bias. A power combiner (PC) can add
a 25ps risetime, 5 ns duration pulse to the pulse pair, and the resulting
signal is then transmitted to the nanomagnet through a 40GHz network
which includes a pick-off tee~(PT) for pulse monitoring {[}Fig.~\ref{fig:fig1}(d),
inset{]} and a second 40 GHz bias tee (BT2) for injecting dc and low
frequency ac currents for device switching and lock-in amplifier measurement
of nanopillar resistance. Device connection is made with a second
40GHz coaxial-to-coplanar probe (P2). For each measurement we reset
the device to the AP state by using a low frequency current ramp,
verify the state of the device by measuring its resistance, send a
pair of ultrafast pulses together with either a 5ns duration pulse
or a dc current, and measure the final state of the device: P or AP.

\begin{figure}
\begin{centering}
\includegraphics[width=3.3in]{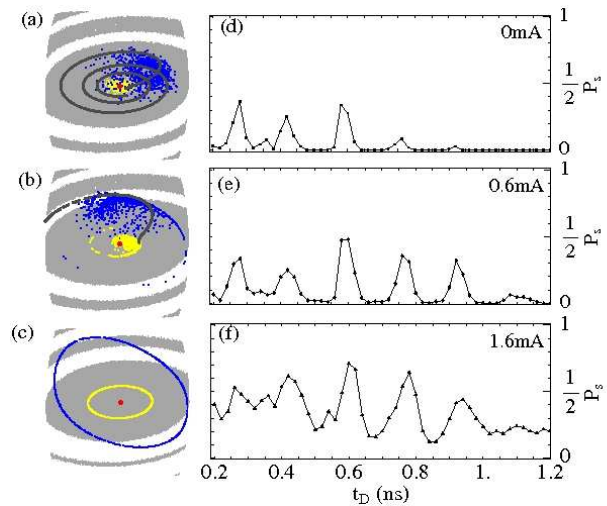}\\
 
\par\end{centering}

\caption{\textbf{dc current effective damping reduction} (a)-(c) Sections of
the free layer moment phase portrait for the dotted region of Fig.~\ref{fig:fig1}(b),
without (a, b) and with (c) dc current. Yellow and blue dots represent
macrospin simulations of a $\mathbf{m}$ ensemble before and after
the (a) first and (b) second ultrafast pulses. The dark line represents
a single trajectory (a) before and (b) after the second ultrafast
pulse. (c) $\mathbf{m}$ ensemble long after applying a dc current
(yellow dots), and same ensemble just after applying the first ultrafast
pulse (blue dots). (d)-(f) $P_{S}$ vs delay between two $\sim$30ps
duration pulses for dc currents of (d) 0 mA, (e) 0.6 mA, and (f) 1.6
mA at room temperature. The {}``dc switching current'', measured
using a sweep rate of 0.05 mA/s is $\sim$1.8mA.}

\label{fig:fig4} 
\end{figure}

The first ultrafast pulse displaces $\mathbf{m}$ away from AP, as
indicated by the dark trajectory simulated in Fig.~\ref{fig:fig4}(a),
and in the absence of spin torque $\mathbf{m}$ relaxes towards AP
due to Gilbert damping. As explained in Ref.~\cite{garzon_PRB2008}
the white and gray regions are the free precession basins of attraction
for P and AP respectively. After some time delay the second ultrafast
pulse displaces $\mathbf{m}$ once again. In the particular case shown
in Fig.~\ref{fig:fig4}(b) $\mathbf{m}$ ends up inside a white band
and thus free evolution of $\mathbf{m}$ leads to switching to the
P state. For certain delays, $\mathbf{m}$ can remain within the first
gray band or might even be excited into large angle precession gray
regions, eventually relaxing towards AP. For every repetition of our
experiment the initial orientation of $\mathbf{m}$ will be different
due to thermal fluctuations, as indicated by the distribution of yellow
dots around the AP state (red) in Fig.~\ref{fig:fig4}(a), where
every point represents a repetition of the experiment. After the first
pulse, the initial $\mathbf{m}$ ensemble shifts to the new one represented
by the blue dots in Fig.~\ref{fig:fig4}(a). Free evolution transforms
it {[}yellow dots in Fig.~\ref{fig:fig4}(b){]}, and the second pulse
pushes the ensemble partially into the first white band {[}blue dots
in Fig.~\ref{fig:fig4}(b){]}, generating a nonzero switching probability
for this particular value of delay. By varying the delay we measure
the switching efficiency of spin torque at different average orientations
of $\mathbf{m}$. In our experiments we measure hundreds of switching
events for each delay to keep the statistical error below 2$\%$.
The bandwidth of our technique, currently 40 GHz, depends on the pulse
duration and can be extended to over 0.8 THz~\cite{nagatsuma_JAP1989,verghese_APL1997}.
Coherent oscillations of $P_{S}$ with delay are shown in Fig.~\ref{fig:fig4}(d).
Here the switching probability peaks, always below P$_{S}$=40$\%$,
decrease dramatically after 600 ps and dissappear before 1000 ps.

Since the nanomagnet effective damping can be modified via STT~\cite{krivorotov_SCIENCE2005,sankey_PRL2006},
we apply dc currents to decrease the effective damping. For 0.6 mA
dc current {[}Fig.~\ref{fig:fig4}(e){]}, we observe an additional
peak at t$_{D}$=1100ps, while the existing peak amplitudes (at 600,
780, and 940 ps) increase, indicating that the dc STT indeed reduces
the damping. As for the zero dc current case, P$_{\text{S}}$ decreases
to 0$\%$ between succesive peaks, demonstrating that a dc current
alone does not contribute to switching. As the dc current increases
to 1.6 mA {[}Fig.~\ref{fig:fig4}(f){]} the peak amplitudes slightly
increase but the peaks broaden substantially. This behavior can be
explained with Fig.~\ref{fig:fig4}(c). At long times after applying
a dc current sufficient to induce continuous precession (i.e. when
the effective damping per cycle vanishes), the $\mathbf{m}$ ensemble
is distributed along a stable orbit (yellow), and thus the resulting
$\mathbf{m}$ ensemble after the first pulse (blue) is largely scattered
in comparison to that obtained in the absence of dc currents. This
scattering not only broadens the $P_{S}$ oscillations but also explains
why $P_{S}$ no longer decreases to 0$\%$ between peaks for 1.6 mA
currents. The dc current together with the first pulse displace some
of the points into the white bands, inducing switching independently
of the second pulse, yielding a nonzero $P_{S}$ baseline.

\begin{figure}
\begin{centering}
\includegraphics[width=3.3in]{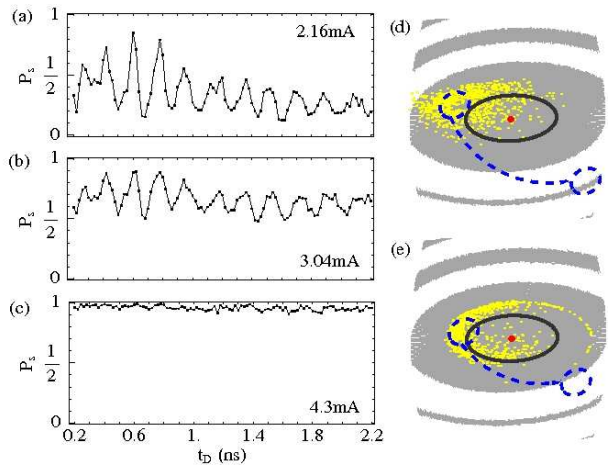}\\
 
\par\end{centering}

\caption{\textbf{5 ns pulse effective damping reduction} (a)-(c) $P_{S}$ vs
delay for 5ns duration pulse amplitudes of (a) 2.16 mA, (b) 3.04 mA,
and (c) 4.30 mA. (d), (e) Sections of the free layer moment phase
portrait for the dotted region of Fig.~\ref{fig:fig1}(b), in the
presence of a 1.2 mA 5ns duration pulsed current. Yellow dots represent
the $\mathbf{m}$ ensemble just before the second ultrafast pulse
for (d) $t_{D}$=140ps and (e) $t_{D}$=580 ps. The solid dark line
represents a stable orbit at 1.2mA while the blue (dashed) regions
schematically represent regimes (i) and (ii) described in the text.}

\label{fig:fig5} 
\end{figure}

To overcome the broadening and dephasing of trajectories in the $\mathbf{m}$
ensemble that occur with dc currents, we employ instead a 25 ps risetime,
5 ns duration current pulse to reduce the effective damping. The first
ultrafast pulse is precisely timed with the beginning of the 5ns duration
pulse throughout our measurements {[}see inset above Fig.~\ref{fig:fig1}(d){]}
and $P_{S}$ is now measured vs. delay for different 5 ns pulse amplitudes,
I$_{5ns}$, as shown in Figs.~\ref{fig:fig5}(a)-(c). In every case
I$_{5ns}$ is greater than the {}``dc'' switching current (measured
using a sweep rate of 0.05 mA/s), and in sharp contrast with Figs.~\ref{fig:fig4}(e)-(f),
Figs.~\ref{fig:fig5}(a)-(b) show P$_{S}$ oscillations persisting
out to our maximum delay, t$_{D}$=2.2 ns. For these 5 ns pulse data
we observe three distinct $P_{S}$ regimes: (i) below 0.6 ns, the
$P_{S}$ maxima are irregular and increase with increasing delay,
(ii) between 0.6-1 ns the $P_{S}$ oscillations monotonically decrease
with increasing delay, and (iii) above 1 ns the oscillation amplitude
is almost constant and shows a minimal decrease at longer $t_{D}$.

These three different regimes can be explained with the phase portrait
sections in Figs.~\ref{fig:fig5}(d) and (e). An initial room temperature
$\mathbf{m}$ distribution equivalent to that in Fig.~\ref{fig:fig4}(a)
evolves to the yellow dots shown in Figs.~\ref{fig:fig5}(d) and
(e) for precession times of 0.14 ns and 0.58 ps respectively in the
presence of a 1.2 mA, 5 ns pulse. At short times {[}Fig.~\ref{fig:fig4}(d){]},
the resulting $\mathbf{m}$ distribution is close to the first gray-white
switching boundary, whereas for longer precession times {[}Fig.~\ref{fig:fig4}(e){]}
the distribution has partially relaxed to the 1.2 mA stable orbit
(dark line). The blue (dashed) ellipses and arrow schematically represent
the evolution of a low temperature $\mathbf{m}$ ensemble in response
to the second ultrafast pulse. For short $t_{D}$ the $\mathbf{m}$
ensemble is displaced across various boundary crossings and the value
of $P_{S}$ depends on the particular phase portrait structure together
with the shape of the thermal distribution, explaining the irregular
behavior of $P_{S}$ at short delays. As $t_{D}$ increases the $\mathbf{m}$
distribution before the second pulse moves towards the 1.2 mA stable
orbit and the $\mathbf{m}$ ensemble resulting from the second pulse
is not pushed as far away from AP. For a particular value of $t_{D}$,
here $\sim$580ps, the $\mathbf{m}$ ensemble is mostly within the
first white band, explaining the increase in $P_{S}$ observed as
$t_{D}$ approaches 600 ps (regime i). As the delay continues to increase
(regime ii), the $\mathbf{m}$ distribution continues to relax towards
the 1.2 mA stable orbit and thus $P_{S}$ decreases monotonically.
This relaxation occurs within 400 ps, a timescale set by the damping.
When the $\mathbf{m}$ distribution has almost reached the stable
orbit, the amplitude of the $P_{S}$ oscillations decreases at a much
slower rate since the decay is now caused by {}``dephasing'' and
not by energy loss (regime iii). Therefore our data are consistent
with complete damping cancellation over a range of currents. The dephasing
of the trajectories corresponding to multiple repetitions of the experiment
and resulting from the thermal distribution of initial orientations
of $\mathbf{m}$ can be observed in Fig.~3(e) (yellow dots). This
dephasing will increase with $t_{D}$ eventually leading to the complete
dephasing shown by the yellow orbit in Fig.~2(c). As the amplitude
of the 5 ns pulse increases, the stable orbit moves away from the
AP stable point and the $\mathbf{m}$ ensemble that results after
the first ultrafast pulse is displaced across the first switching
boundary, producing the background $P_{S}$ observed in our measurements.

In our simulations and discussion we have ignored thermal effects
on the magnetic moment motion. Thermal fluctuations will introduce
a stochastic component into the magnetization trajectories, broadening
the $\mathbf{m}$ distribution further~\cite{brown_PHYSREV1963}.
This broadening can also contribute to the existence of a nonzero
$P_{S}$ background, and for long delays it will produce additional
dephasing of the trajectories and eventually a dissappearance of the
$P_{S}$ oscillations. By extending our technique to much longer times
it should be possible to observe the dephasing of the ensemble of
trajectories even though each trajectory on average does not lose
any energy per precession cycle.

In conclusion, we demonstrate time domain detection of damping modification
in a nanomagnet at room temperature, and in contrast to dc currents,
successfully achieve damping cancellation with a precisely-timed 5
ns pulse over a range of currents. In addition, our ultrafast time-domain
measurement allows us to monitor the transient approach to an equilibrium
precessional trajectory, where the damping cancellation and linewidth
reduction seen in spin torque oscillators are established after a
transient delay while the magnetization decays to the equilibrium
orbit. These measurements suggest the possibility to further study
thermal dephasing effects in oscillators, both while relaxing towards
stable precession, and for times much longer than the precession period.

\bibliography{references}

\begin{thebibliography}{15}
\expandafter\ifx\csname natexlab\endcsname\relax\def\natexlab#1{#1}\fi
\expandafter\ifx\csname bibnamefont\endcsname\relax
  \def\bibnamefont#1{#1}\fi
\expandafter\ifx\csname bibfnamefont\endcsname\relax
  \def\bibfnamefont#1{#1}\fi
\expandafter\ifx\csname citenamefont\endcsname\relax
  \def\citenamefont#1{#1}\fi
\expandafter\ifx\csname url\endcsname\relax
  \def\url#1{\texttt{#1}}\fi
\expandafter\ifx\csname urlprefix\endcsname\relax\def\urlprefix{URL }\fi
\providecommand{\bibinfo}[2]{#2}
\providecommand{\eprint}[2][]{\url{#2}}

\bibitem[{\citenamefont{{Slonczewski}}(1996)}]{slonczewski_JMMM1996}
\bibinfo{author}{\bibfnamefont{J.~C.} \bibnamefont{{Slonczewski}}},
  \bibinfo{journal}{J. Magn. Magn. Mater.} \textbf{\bibinfo{volume}{159}},
  \bibinfo{pages}{L1} (\bibinfo{year}{1996}).

\bibitem[{\citenamefont{Berger}(1996)}]{berger_PRB1996}
\bibinfo{author}{\bibfnamefont{L.}~\bibnamefont{Berger}},
  \bibinfo{journal}{Phys. Rev. B} \textbf{\bibinfo{volume}{54}},
  \bibinfo{pages}{9353} (\bibinfo{year}{1996}).

\bibitem[{\citenamefont{Ralph and Stiles}(2008)}]{ralph_JMMM2007}
\bibinfo{author}{\bibfnamefont{D.}~\bibnamefont{Ralph}} \bibnamefont{and}
  \bibinfo{author}{\bibfnamefont{M.}~\bibnamefont{Stiles}},
  \bibinfo{journal}{Journal of Magnetism and Magnetic Materials}
  \textbf{\bibinfo{volume}{320}}, \bibinfo{pages}{1190 }
  (\bibinfo{year}{2008}).

\bibitem[{\citenamefont{Krivorotov et~al.}(2005)\citenamefont{Krivorotov,
  Emley, Sankey, Kiselev, Ralph, and Buhrman}}]{krivorotov_SCIENCE2005}
\bibinfo{author}{\bibfnamefont{I.~N.} \bibnamefont{Krivorotov}},
  \bibinfo{author}{\bibfnamefont{N.~C.} \bibnamefont{Emley}},
  \bibinfo{author}{\bibfnamefont{J.~C.} \bibnamefont{Sankey}},
  \bibinfo{author}{\bibfnamefont{S.~I.} \bibnamefont{Kiselev}},
  \bibinfo{author}{\bibfnamefont{D.~C.} \bibnamefont{Ralph}}, \bibnamefont{and}
  \bibinfo{author}{\bibfnamefont{R.~A.} \bibnamefont{Buhrman}},
  \bibinfo{journal}{Science} \textbf{\bibinfo{volume}{307}},
  \bibinfo{pages}{228} (\bibinfo{year}{2005}).

\bibitem[{\citenamefont{Sankey et~al.}(2006)\citenamefont{Sankey, Braganca,
  Garcia, Krivorotov, Buhrman, and Ralph}}]{sankey_PRL2006}
\bibinfo{author}{\bibfnamefont{J.~C.} \bibnamefont{Sankey}},
  \bibinfo{author}{\bibfnamefont{P.~M.} \bibnamefont{Braganca}},
  \bibinfo{author}{\bibfnamefont{A.~G.~F.} \bibnamefont{Garcia}},
  \bibinfo{author}{\bibfnamefont{I.~N.} \bibnamefont{Krivorotov}},
  \bibinfo{author}{\bibfnamefont{R.~A.} \bibnamefont{Buhrman}},
  \bibnamefont{and} \bibinfo{author}{\bibfnamefont{D.~C.} \bibnamefont{Ralph}},
  \bibinfo{journal}{Physical Review Letters} \textbf{\bibinfo{volume}{96}},
  \bibinfo{pages}{227601} (\bibinfo{year}{2006}).

\bibitem[{\citenamefont{Tsoi et~al.}(1998)\citenamefont{Tsoi, Jansen, Bass,
  Chiang, Seck, Tsoi, and Wyder}}]{tsoi_PRL1998}
\bibinfo{author}{\bibfnamefont{M.}~\bibnamefont{Tsoi}},
  \bibinfo{author}{\bibfnamefont{A.~G.~M.} \bibnamefont{Jansen}},
  \bibinfo{author}{\bibfnamefont{J.}~\bibnamefont{Bass}},
  \bibinfo{author}{\bibfnamefont{W.-C.} \bibnamefont{Chiang}},
  \bibinfo{author}{\bibfnamefont{M.}~\bibnamefont{Seck}},
  \bibinfo{author}{\bibfnamefont{V.}~\bibnamefont{Tsoi}}, \bibnamefont{and}
  \bibinfo{author}{\bibfnamefont{P.}~\bibnamefont{Wyder}},
  \bibinfo{journal}{Phys. Rev. Lett.} \textbf{\bibinfo{volume}{80}},
  \bibinfo{pages}{4281} (\bibinfo{year}{1998}).

\bibitem[{\citenamefont{Katine et~al.}(2000)\citenamefont{Katine, Albert,
  Buhrman, Myers, and Ralph}}]{katine_PRL2000}
\bibinfo{author}{\bibfnamefont{J.~A.} \bibnamefont{Katine}},
  \bibinfo{author}{\bibfnamefont{F.~J.} \bibnamefont{Albert}},
  \bibinfo{author}{\bibfnamefont{R.~A.} \bibnamefont{Buhrman}},
  \bibinfo{author}{\bibfnamefont{E.~B.} \bibnamefont{Myers}}, \bibnamefont{and}
  \bibinfo{author}{\bibfnamefont{D.~C.} \bibnamefont{Ralph}},
  \bibinfo{journal}{Phys. Rev. Lett.} \textbf{\bibinfo{volume}{84}},
  \bibinfo{pages}{3149} (\bibinfo{year}{2000}).

\bibitem[{\citenamefont{Sankey et~al.}(2005)\citenamefont{Sankey, Krivorotov,
  Kiselev, Braganca, Emley, Buhrman, and Ralph}}]{sankey_PRB2005}
\bibinfo{author}{\bibfnamefont{J.~C.} \bibnamefont{Sankey}},
  \bibinfo{author}{\bibfnamefont{I.~N.} \bibnamefont{Krivorotov}},
  \bibinfo{author}{\bibfnamefont{S.~I.} \bibnamefont{Kiselev}},
  \bibinfo{author}{\bibfnamefont{P.~M.} \bibnamefont{Braganca}},
  \bibinfo{author}{\bibfnamefont{N.~C.} \bibnamefont{Emley}},
  \bibinfo{author}{\bibfnamefont{R.~A.} \bibnamefont{Buhrman}},
  \bibnamefont{and} \bibinfo{author}{\bibfnamefont{D.~C.} \bibnamefont{Ralph}},
  \bibinfo{journal}{Phys. Rev. B} \textbf{\bibinfo{volume}{72}},
  \bibinfo{pages}{224427} (\bibinfo{year}{2005}).

\bibitem[{\citenamefont{Houssameddine et~al.}(2009)\citenamefont{Houssameddine,
  Ebels, Dieny, Garello, Michel, Delaet, Viala, Cyrille, Katine, and
  Mauri}}]{houssameddine_PRL2009}
\bibinfo{author}{\bibfnamefont{D.}~\bibnamefont{Houssameddine}},
  \bibinfo{author}{\bibfnamefont{U.}~\bibnamefont{Ebels}},
  \bibinfo{author}{\bibfnamefont{B.}~\bibnamefont{Dieny}},
  \bibinfo{author}{\bibfnamefont{K.}~\bibnamefont{Garello}},
  \bibinfo{author}{\bibfnamefont{J.-P.} \bibnamefont{Michel}},
  \bibinfo{author}{\bibfnamefont{B.}~\bibnamefont{Delaet}},
  \bibinfo{author}{\bibfnamefont{B.}~\bibnamefont{Viala}},
  \bibinfo{author}{\bibfnamefont{M.-C.} \bibnamefont{Cyrille}},
  \bibinfo{author}{\bibfnamefont{J.~A.} \bibnamefont{Katine}},
  \bibnamefont{and} \bibinfo{author}{\bibfnamefont{D.}~\bibnamefont{Mauri}},
  \bibinfo{journal}{Physical Review Letters} \textbf{\bibinfo{volume}{102}},
  \bibinfo{eid}{257202} (pages~\bibinfo{numpages}{4}) (\bibinfo{year}{2009}).

\bibitem[{\citenamefont{Katine and Fullerton}(2008)}]{katine_JMMM2008}
\bibinfo{author}{\bibfnamefont{J.}~\bibnamefont{Katine}} \bibnamefont{and}
  \bibinfo{author}{\bibfnamefont{E.~E.} \bibnamefont{Fullerton}},
  \bibinfo{journal}{Journal of Magnetism and Magnetic Materials}
  \textbf{\bibinfo{volume}{320}}, \bibinfo{pages}{1217 }
  (\bibinfo{year}{2008}), ISSN \bibinfo{issn}{0304-8853}.

\bibitem[{\citenamefont{Garzon et~al.}(2008)\citenamefont{Garzon, Ye, Webb,
  Crawford, Covington, and Kaka}}]{garzon_PRB2008}
\bibinfo{author}{\bibfnamefont{S.}~\bibnamefont{Garzon}},
  \bibinfo{author}{\bibfnamefont{L.}~\bibnamefont{Ye}},
  \bibinfo{author}{\bibfnamefont{R.~A.} \bibnamefont{Webb}},
  \bibinfo{author}{\bibfnamefont{T.~M.} \bibnamefont{Crawford}},
  \bibinfo{author}{\bibfnamefont{M.}~\bibnamefont{Covington}},
  \bibnamefont{and} \bibinfo{author}{\bibfnamefont{S.}~\bibnamefont{Kaka}},
  \bibinfo{journal}{Physical Review B (Condensed Matter and Materials Physics)}
  \textbf{\bibinfo{volume}{78}}, \bibinfo{pages}{180401}
  (\bibinfo{year}{2008}).

\bibitem[{\citenamefont{Auston}(1984)}]{auston_1984}
\bibinfo{author}{\bibfnamefont{D.~H.} \bibnamefont{Auston}}, in
  \emph{\bibinfo{booktitle}{Picosecond Optoelectronic Devices}}, edited by
  \bibinfo{editor}{\bibfnamefont{C.~H.} \bibnamefont{Lee}}
  (\bibinfo{publisher}{Academic, Orlando}, \bibinfo{year}{1984}), pp.
  \bibinfo{pages}{73--117}.

\bibitem[{\citenamefont{Nagatsuma et~al.}(1989)\citenamefont{Nagatsuma,
  Shibata, Sano, and Iwata}}]{nagatsuma_JAP1989}
\bibinfo{author}{\bibfnamefont{T.}~\bibnamefont{Nagatsuma}},
  \bibinfo{author}{\bibfnamefont{T.}~\bibnamefont{Shibata}},
  \bibinfo{author}{\bibfnamefont{E.}~\bibnamefont{Sano}}, \bibnamefont{and}
  \bibinfo{author}{\bibfnamefont{A.}~\bibnamefont{Iwata}},
  \bibinfo{journal}{Journal of Applied Physics} \textbf{\bibinfo{volume}{66}},
  \bibinfo{pages}{4001} (\bibinfo{year}{1989}).

\bibitem[{\citenamefont{Verghese et~al.}(1997)\citenamefont{Verghese, Zamdmer,
  Hu, and F{\"o}rster}}]{verghese_APL1997}
\bibinfo{author}{\bibfnamefont{S.}~\bibnamefont{Verghese}},
  \bibinfo{author}{\bibfnamefont{N.}~\bibnamefont{Zamdmer}},
  \bibinfo{author}{\bibfnamefont{Q.}~\bibnamefont{Hu}}, \bibnamefont{and}
  \bibinfo{author}{\bibfnamefont{A.}~\bibnamefont{F{\"o}rster}},
  \bibinfo{journal}{Applied Physics Letters} \textbf{\bibinfo{volume}{70}},
  \bibinfo{pages}{2644} (\bibinfo{year}{1997}).

\bibitem[{\citenamefont{Brown}(1963)}]{brown_PHYSREV1963}
\bibinfo{author}{\bibfnamefont{W.~F.} \bibnamefont{Brown}},
  \bibinfo{journal}{Phys. Rev.} \textbf{\bibinfo{volume}{130}},
  \bibinfo{pages}{1677} (\bibinfo{year}{1963}).

\end{thebibliography}

\end{document}